\documentclass[11pt,a4]{article}
\pdfoutput=1

\usepackage{jheppub} 
\usepackage[utf8]{inputenc}
\usepackage{amsfonts}
\usepackage{amsbsy}
\usepackage{url}
\usepackage{enumerate}
\usepackage[font=small,justification=justified]{caption}
\usepackage{hhline}
\usepackage{graphicx,color}
\usepackage{amssymb,amsmath}
\usepackage{slashed}
\usepackage{hyperref}
\usepackage{braket}
\usepackage{simplewick}
\usepackage{latexsym}
\usepackage{pifont}          
\usepackage{bm}
\usepackage{comment}
\usepackage[normalem]{ulem}
\usepackage{cancel}
\usepackage{amsmath}
\usepackage{amssymb}
\usepackage{nccmath}
\usepackage{graphicx}
\usepackage{here}
\usepackage{subcaption}
\usepackage{url}
\usepackage{braket}
\usepackage{physics}% Added by SK
\usepackage[compat=1.1.0]{tikz-feynhand}
\usepackage{slashed}
\usepackage{mathtools}
\usepackage{bbold}
\usepackage{todonotes}
\usepackage{orcidlink}

\graphicspath{{./fig/}{./}{./figs/}}

\newcommand{\mr}[1]{\mathrm{#1}}

\newcommand{\1}{\mathbb{1}}
\newcommand{\alias}[3]{%
  \let#3=#2
  \newcommand{#1}{#3}
  \renewcommand{#2}{{\textcolor{red}{#3}}}
}
\newcommand{\Vlambda}{\tilde{\lambda}}
\newcommand{\Hlambda}{\lambda}

\newcommand{\rhoE}{\rho_{\mathrm{E}}}

\newcommand{\tE}{t_{\mathrm{E}}}

\newcommand{\brad}{{\rhoE^*}}

\newcommand{\SU}{\mathrm{SU}}
\newcommand{\pauli}[1]{\sigma^{#1}}

\numberwithin{equation}{section}

\begin{document}
\def\ps{\mathbf{p}}
\def\PS{\mathbf{P}}
\baselineskip 0.6cm
\def\simgt{\mathrel{\lower2.5pt\vbox{\lineskip=0pt\baselineskip=0pt
           \hbox{$>$}\hbox{$\sim$}}}}
\def\simlt{\mathrel{\lower2.5pt\vbox{\lineskip=0pt\baselineskip=0pt
           \hbox{$<$}\hbox{$\sim$}}}}
\def\simprop{\mathrel{\lower3.0pt\vbox{\lineskip=1.0pt\baselineskip=0pt
             \hbox{$\propto$}\hbox{$\sim$}}}}
\def\tr{\mathop{\rm tr}}
\def\SU{\mathop{\rm SU}}

\def\azimuth{\varphi}
\newcommand{\az}{\azimuth}
\def\zenith{\theta}
\def\Higgsprofile{\xi}
\def\Newton{G_N}

\preprint{CERN-TH-2026-106, OU-HET-1315}

\title{
The decay rate of metastable cosmic strings \\ beyond the thin-string approximation
}

\author[a]{Valerie~Domcke~\orcidlink{0000-0002-7208-4464}}
\emailAdd{valerie.domcke@cern.ch}
\affiliation[a]{Theoretical Physics Department, CERN, 1 Esplanade des Particules, CH-1211 Geneva 23, Switzerland}

\author[b]{and Yu~Hamada~\orcidlink{0000-0002-0227-5919}}
\emailAdd{
%yu.hamada@desy.de
yu.hamada@het.phys.sci.osaka-u.ac.jp
}
\affiliation[b]{Department of Physics, The University of Osaka,
Machikaneyama-cho, Toyonaka, Osaka 560- 0043, Japan}
%\affiliation[c]{Deutsches Elektronen-Synchrotron DESY, Notkestr.~85, 22607 Hamburg, Germany}
%\affiliation[d]{Research and Education Center for Natural Sciences, Keio University, 4-1-1 Hiyoshi, Yokohama, Kanagawa 223-8521, Japan}

\abstract{In the context of grand unified theories, any cosmic strings present in the post-inflationary universe are likely to be metastable, with a decay rate set by the spontaneous creation of monopole pairs on the string. Determining this decay rate is crucial in understanding the phenomenology of the cosmic string network, including a potentially observable gravitational wave background. The bounce action governing this rate has so far only been determined using the thin string approximation or specific ans\"atze for the field profiles in the monopole formation process. Here we solve this problem using classical lattice simulations, relying only on the inherent symmetries of the problem. Our results indicate a suppression of the bounce action and hence a faster string decay compared to previous estimates.}

\maketitle

%%%%%%%%%%%%%%%%%%%%
\section{Introduction}

Cosmic strings are formed in spontaneous symmetry breaking processes in the early universe, $G \mapsto H$, if the vacuum manifold of the quotient group features a non-trivial first homotopy class, $\pi_1(G/H) \neq 1$~\cite{Kibble:1976sj,Vilenkin:2000jqa}. The simplest example is the spontaneous breaking of a $U(1)$ symmetry. Once formed, the cosmic string network evolves to reach the scaling regime, in which a constant fraction of the energy budget of the Universe is maintained in the cosmic string network. This is achieved by the (self-)intersection of cosmic strings, which leads to the formation of cosmic string loops which can decay through the emission of gravitational waves (GWs). Searches for stochastic gravitational wave backgrounds (SGWBs)~\cite{LIGOScientific:2025kry,NANOGrav:2023hvm} and cosmic microwave background (CMB) observations~\cite{Planck:2015fie} place \textit{upper} bounds on the cosmic string tension and hence the symmetry breaking scale of about $10^{15}$~GeV.
%more details: LIGO, conservative prior, gives $G\mu < 3.7 \cdot 10^{-8} \rightarrow v < 9.4 \cdot 10^{14}$~GeV. CMB gives $G\mu < 10^{-7} \rightarrow v < 1.5 \cdot 10^{15}$~GeV.

In the context of grand unified theories (GUTs), several symmetry breaking steps are often required to reach the Standard Model. This can lead to the formation of metastable cosmic strings: A first phase transition leading to monopoles is followed by an inflationary phase which dilutes the monopoles to the point of eliminating them from the observable Universe. A second phase transition leads to the formation of cosmic strings. The presence of the monopoles in the theory allows for the spontaneous creation of monopole antimonopole pairs along the string with a rate~\cite{Preskill:1992ck,Monin:2008mp,Leblond:2009fq}
\begin{align}
  \Gamma \sim \frac{\mu}{2 \pi} e^{- \pi m_M^2/\mu} \,,
  \label{eq:Gamma_intro}
\end{align}
per unit length of string. Here, $m_M$ is the monopole mass and $\mu$ the string tension. This breaks the strings and leads to a decay of the cosmic string network. If $m_M^2 \gg \mu$, the decay rate is suppressed and the cosmic string lifetime exceeds the age of the universe, recovering the usual phenomenology of stable cosmic strings. If the symmetry breaking scales are less hierarchical, $m_M^2 \gtrsim \mu$, the strings may not live long enough to imprint on the CMB (and are as such not subject to the corresponding bound~\cite{Planck:2015fie}) and will lead to a GW spectrum that is suppressed at low frequencies~\cite{Buchmuller:2021mbb}, as GWs with low frequencies are generated only at late times. This is particularly interesting for GW phenomenology, as the SGWB produced from a metastable cosmic string network provides a remarkably good fit to the tentative signal observed at pulsar timing arrays~\cite{NANOGrav:2023hvm} and opens up the possibility of a detection at LIGO/Virgo/KAGRA (LVK) or at third generation ground-based interferometers~\cite{Buchmuller:2020lbh}.

However, in the near degenerate case $m_M^2 \gtrsim \mu$ of particular interest, Eq.~\eqref{eq:Gamma_intro} is expected to provide a particularly poor approximation. The derivation is based on the thin string approximation with $m_M$ the mass of an isolated monopole in vacuum and $\mu$ the tension of a stable cosmic string. In the absence of any scale separation between the width of the string, the radius of the monopole, and separation of the monopole antimonopole pair at nucleation, this treatment is no longer justified. The goal of this paper is to provide an accurate expression for Eq.~\eqref{eq:Gamma_intro} which overcomes this shortcoming. From the discussion above it is clear that this will crucially impact the expected phenomenology.

Our work builds in particular on Ref.~\cite{Chitose:2023dam}, which reformulated the task as the computation of the bounce action associated with the formation of a monopole in a $SU(2)$ gauge theory. Using different ans\"atze (in particular the one proposed in~\cite{Shifman:2002yi}) for the Higgs and gauge field configurations, the authors derived an upper bound for the bounce action, which provides a lower bound for the decay rate. Our work goes beyond this by generalizing to the most general symmetry-restricted ans\"atz which allows us to recover the thin string limit~\eqref{eq:Gamma_intro} as we increase the scale separation. This confirms that the solution we find for the field configurations is indeed the minimal energy one, so that our results are not only an upper bound but provide the actual value of the decay rate.

Our results show a significant suppression of the bounce action (and hence an increase in the decay rate) as we move from hierarchical to near degenerate symmetry breaking scales. This shifts the phenomenologically most interesting parameter space to somewhat larger values %($\sim 10$)
of the ratio of symmetry breaking scales and introduces a new dependence on the $SU(2)$ gauge coupling. The former facilitates the implementation into a GUT framework, as achieving two distinct symmetry breaking steps with an intermediate phase of inflation becomes easier.

The remainder of the paper is structured as follows. Section~\ref{sec:setup} introduces our setup, which is based on a two-step breaking of an $SU(2)$ gauge symmetry. We revisit the properties of isolated strings and monopoles in this framework and derivation of the decay rate in the thin string limit. We then in Sec.~\ref{sec:numerics} turn to the proposed method of computing the bounce action going beyond the thin string approximation. We discuss symmetries of gauge and Higgs field configurations. Lattice simulations using gradient flow allow us to numerically determine the saddle points of the bounce action and infer the string decay rate. Our results and implications for GW searches are presented in Sec.~\ref{sec:results} before concluding in Sec.~\ref{sec:conclusions}.

Four appendices provide technical details. Appendix~\ref{app:convention} explains the dimensionless variables used in our numerical simulations. Appendix~\ref{app:classical-stability} discusses the classical (in)stability of metastable cosmic strings. Appendix~\ref{app:mass-ratio} provides results for parameter choices complementary to those discussed in the main text. In App.~\ref{app:SGWB} we review the GW spectrum from metastable cosmic strings in the thin string approximation.

%%%%%%%%%%%%%%%%%%%%%%%%%%%%%%%%%%%%%%%%%%%%%%%%%%%%%%%%%%%%%%%%%%%%%%%%%%%%%%%%%%%%%%%%%%%%%%%%%%%%%%%%%%%%%%

\section{A minimal setup for metastable cosmic strings}
\label{sec:setup}
We first specify a minimal framework in which metastable cosmic strings can form, closely following the notation of Ref.~\cite{Chitose:2023dam}. The starting point is an $SU(2)$ gauge theory with two successive stages of symmetry breaking. These are controlled by an adjoint scalar field $\phi^a$ ($a=1,2,3$) and a doublet scalar field $H_i$ ($i=1,2$),
\begin{equation} \label{eq:lagrangian}
\mathcal{L}=\frac{1}{2}(D_\mu \phi^a)(D^\mu \phi^a) + (D_\mu H)^\dagger (D_\mu H) - \frac{1}{4g^2}F^a_{\mu\nu}F^{a\mu\nu}-V_\mathrm{Higgs}(\phi,H)\ .
\end{equation}
Here, $F_{\mu\nu}^a$ is the field strength tensor, $g$ denotes the gauge coupling, and the scalar covariant derivatives are defined as
\begin{align}
    D_\mu H\equiv \partial_\mu H  - i \frac{\sigma^{a}}{2}A_\mu^a H \ ,
    \qquad
    D_\mu \phi^a \equiv \partial_\mu\phi^a + \epsilon^{abc}A_\mu^b \phi^c\ ,
\end{align}
where the $\pauli{}$'s are the Pauli matrices and the doublet indices are suppressed.
The scalar potential is chosen to realize the required symmetry breaking pattern,
\begin{equation}
    V_{\mathrm{Higgs}}(\phi, H )\equiv \Hlambda\pqty{\abs{ H }^2-v^2}^2+\Vlambda\pqty{\phi^a\phi^a-V^2}^2+\gamma \abs{\pqty{\phi-\frac{V}{2} \mathbb{1}} H }^2\ ,
    \label{eq:VHiggs}
\end{equation}
where the dimensionless couplings $\Hlambda$, $\Vlambda$ and $\gamma$ are assumed to be positive, and we have introduced $\phi\equiv\phi^a\pauli{a}/2$. We focus on the hierarchy $V > v$, so that monopoles are produced before cosmic strings. The first symmetry breaking step, together with monopole formation, takes place at the scale $V$, when $\phi^a$ develops the vacuum expectation value (vev),
\begin{equation}
\label{eq:VEV1}
   \langle{\phi^a}\rangle = V \delta^{a3} \,.
\end{equation}
Using $SU(2)$ gauge invariance, we have oriented the vev along the $a=3$ direction without loss of generality. This reduces the $SU(2)$ symmetry to its diagonal $U(1)$ subgroup, corresponding to the $SO(2)$ rotations about the $a=3$ axis of $SU(2)$.
In this phase, two components of the adjoint Higgs $\phi^a$ become the longitudinal modes of two $SU(2)$ gauge bosons with mass $m_W$. The remaining spectrum contains one real scalar of mass $m_\phi$ and a massless gauge boson associated with the unbroken $U(1)$ symmetry,
\begin{align}
    m_\phi = \sqrt{8\tilde{\lambda}} V \ , \quad m_W = g V \,.
\end{align}
In addition, the last term in Eq.~\eqref{eq:VHiggs} (where we have suppressed the doublet indices $i$) gives a large mass to the lower component of the Higgs doublet $H$ as $m_{h_2}=\sqrt{\gamma}\,V$. The symmetry breaking driven by $\phi^3$ at the scale $V$ produces monopoles, since the second homotopy group is non-trivial, $\pi_2(SU(2)/U(1))\simeq \mathbb{Z}$.

At the second stage, the upper component of the scalar doublet $H$ acquires a vev at the scale $v$,
\begin{align}
    \langle H\rangle =
\begin{pmatrix}
v \\0
\end{pmatrix}
\end{align}
thereby breaking the remaining $U(1)$ symmetry and giving rise to cosmic strings. Since the $U(1)$ charge of $H$ is $\pm 1/2$, this also breaks the $\mathbb{Z}_2$ center symmetry of $SU(2)$. The masses of the remaining gauge boson and Higgs field are
\begin{align}
 m_\gamma = \frac{g}{\sqrt{2}} v \,,  \quad  m_{ h _1} = 2\sqrt{\Hlambda} v\,.
\end{align}

We do not specify here a concrete dynamical model that realizes these successive symmetry breaking steps. The scenario we have in mind (see e.g.~\cite{Buchmuller:2023aus,Buchmuller:2021dtt,Antusch:2023zjk,Maji:2025thf}) is a hybrid inflation scenario.
%consists of two stages of symmetry breaking separated by a period of inflation. 
For instance, a slowly rolling scalar field causes the symmetry breaking leading to the monopole production and drives inflation to dilute them in the first step.
Then it experiences the tachyonic instability at the end of inflation and triggers the $U(1)$ breaking, forming the cosmic strings in the second step.
Those strings are stable except for the non-perturbative nucleation of a monopole--antimonopole pair, which can ``break'' the string. 
This differs from the scenario analyzed in Ref.~\cite{Tranchedone:2026lav}, where the symmetry breaking transitions are driven by thermal or quantum fluctuations.

\subsection{Monopoles}
For hierarchical scales $V \gg v$, monopole and string formation can be treated as separate processes. We first recall this analytically tractable limit, before turning to a numerical study of nearly degenerate symmetry breaking scales.

As discussed above, the first breaking step $SU(2)\to U(1)$ gives rise to 't Hooft-Polyakov monopoles~\cite{tHooft:1974kcl,Polyakov:1974ek}. A static monopole with unit winding number, centered at the origin, is described by
\begin{align}
\label{eq:heg}
\phi^a=V G(r)\displaystyle{\frac{x^a}{r}}\ , \quad
A_{0}^a = 0 \ ,\quad
A^{a}_{i}=
\displaystyle{\frac{\epsilon^{aij}x^j}{r^2}}F(r)\ ,\quad (i,j=1,2,3)\ ,
\end{align}
where
$r\equiv\sqrt{x^2+y^2+z^2}$.
The profile functions $G(r)$ and $F(r)$ obey the boundary conditions
\begin{equation}
G(r)\to 0 , \quad F(r)\to 0  \quad (r\to 0) \,, \label{eq:asymp-mon-r0}
\end{equation}
\begin{equation}
G(r)\to 1 , \quad F(r)\to 1  \quad (r\to \infty) \,, \label{eq:asymp-mon-rInf}
\end{equation}
and approach their asymptotic values exponentially as $r\to \infty$.\footnote{
Here and in the following, we will not consider the Bogomol'nyi–Prasad–Sommerfield limit $\tilde{\lambda}\to0$ for which $m_\phi \to 0$.
}

It is useful to introduce the gauge-invariant field strength of the unbroken $U(1)$ gauge field,
\begin{equation}
\label{eq:effectiveF}
F^{U(1)}_{\mu\nu} \equiv \frac{1}{V}\phi^a F^{a}_{\mu\nu}
\end{equation}
together with the corresponding magnetic field
\begin{equation}
B^{U(1)}{}^i \equiv \frac{\varepsilon^{ijk}}{2}F^{U(1)}_{jk} \, . \label{eq:U(1)flux}
\end{equation}
Using the asymptotic behaviour at large distances in Eq.~\eqref{eq:asymp-mon-rInf}, one finds that the monopole magnetic field has a hedgehog form,
\begin{equation}
B^{U(1)}{}^i  \to -
\frac{x^i}{r^3}\, .
\end{equation}
The corresponding magnetic charge is therefore
\begin{equation}
\label{eq: monopole flux}
    Q^m \equiv \int_{r \to \infty}
    dS_{i}\, B^{U(1)}{}^{i  } =  -4\pi \ ,
\end{equation}
where $d{S}_{i}$ denotes the surface element of a two-dimensional sphere surrounding the monopole.

The monopole mass $m_M$ may be written as~\cite{Bogomolny:1976ab,Kirkman:1981ck}
\begin{align}
    m_M = c_M \, \frac{4\pi V}{g}  \, ,
\end{align}
where $c_M$ is an order one factor determined by the particle masses $m_\phi$ and $m_W$. For the benchmark used below, $m_\phi = m_W$, one has $c_M = 1.23768$~\cite{Forgacs:2005vx}. More generally, the Bogomolny bound implies $c_M \geq 1$.

\subsection{Cosmic strings}
\label{sec:CosmicString}
We now consider the Abrikosov-Nielsen-Olesen (ANO) strings~\cite{Abrikosov:1956sx,Nielsen:1973cs} produced during the second symmetry breaking step. In the decoupling limit, we take the vacuum expectation value of $\phi$ in Eq.~\eqref{eq:VEV1} as fixed, leading to $F_{\mu\nu}^{U(1)} = F_{\mu\nu}^3$. All particles with masses proportional to $V$ then decouple, leaving the $U(1)$ sector containing the gauge field $A^3$ and the complex scalar field $h_1$.
The covariant derivative is
\begin{align}
    D_\mu h_1 = \left(\partial_\mu -  \frac{i}{2}A_\mu^3\right)h_1 \ .
\end{align}
The static string solution along the $x^3$-axis takes the form (see e.g., Ref.~\cite{Vilenkin:2000jqa})
\begin{align}
\label{eq:string ansatz1}
%H(\rho) &= \mqty(h_1(\rho)\\0)\  ,\\
h_1(\rho) &=v\Higgsprofile(\rho)e^{-i n\azimuth}\ , \\
\label{eq:string ansatz2}
A_{i}^3&={2n}\frac{\epsilon_{ij}x^j}{\rho^2}(1-f(\rho))\ ,~~~~(i,j=1,2)\ , \\
A^3_{0} &= A^3_{3} = 0 \ ,
\end{align}
where $n \in \mathbb{Z}$ is the string winding number, and $\Higgsprofile(\rho)$ and $f(\rho)$ are profile functions.
We use cylindrical coordinates, $\azimuth\equiv{\arctan}(x_2/x_1)$ and $\rho\equiv\sqrt{x_1^2+x_2^2}$.
The two-dimensional antisymmetric tensor is normalized as $\epsilon_{12}=1$.
The profile functions satisfy
\begin{align}
\Higgsprofile(\rho)\rightarrow 0\ , ~(\rho\rightarrow0)&\ ,~~~~~\Higgsprofile(\rho)\rightarrow 1\ , ~(\rho\rightarrow \infty)\ ,\\
f(\rho)\rightarrow 1\ , ~(\rho\rightarrow 0)&\ ,~~~~~f(\rho)\rightarrow 0\ ,
~(\rho\rightarrow \infty)\ .
\end{align}
They approach their asymptotic values exponentially for $\rho \gg (gv)^{-1}, (\sqrt{\lambda}v)^{-1}$. Moreover, $D_\mu h_1$ vanishes exponentially in this limit, ensuring a finite string tension.

The winding number determines the magnetic flux carried by the string,
\begin{align}
\label{eq: string flux}
\int \dd[2]{x}B_{3}^{U(1)}=\oint_{\rho\rightarrow \infty} A_{i}^3 \dd{x}^i=-{4\pi n}\ .
\end{align}
Thus, for $|n|=1$, the magnetic flux along the string matches the magnetic charge of a magnetic (anti)monopole.

The string tension can be parameterized as~\cite{Hill:1987ye,deVega:1976xbp}
\begin{align}
    \mu = c_S \, 2\pi v^2\,,
\end{align}
where $c_S \sim 1$ is an order one constant depending on the gauge boson and Higgs masses, $m_\gamma$ and $m_{h_1}$. In the special case used below, $m_\gamma = m_{h_1}$, one has $c_S = 1$.

%%%%%%%%%%%%%%%%%%%%
\subsection{Breaking of infinitely thin strings}
\label{sec: Thin String}

Here, we briefly review the conventional estimate of the string breaking rate in the infinitely thin cosmic string limit~\cite{Preskill:1992ck}.
As discussed above, cosmic strings can decay by spontaneous creation of monopole antimonopole pairs, which we can describe as a tunneling process.

To calculate the tunneling factor, we use the Euclidean path integral method ($t = -i\tE$).
In the infinitely thin string limit, the string breaking process may be regarded as a vacuum decay in 1+1 dimensions. Taking the $z$-axis to be aligned with the cosmic string, the presence of the cosmic string along this axis corresponds to the false vacuum, and the absence of string corresponds to the true vacuum. The monopole plays the role of the domain wall separating the two vacua.

In Minkowski space, the cosmic string is invariant with respect to Lorentz boosts along the string, i.e.\ Lorentz boosts in the $(t,z)$ plane, which correspond to rotations in the Euclidean $(\tE,z)$ plane.
We assume that the bounce solution preserves this symmetry, and hence, 
the domain wall separating the two vacua (i.e. the monopole worldline) is a circle on
the $(\tE,z)$ plane.

The bounce action of the bubble is given by,
\begin{align}
    S_B&=m_M\int_{\text{worldline}} \dd{x}- \mu\int_{\text{hole}}\dd[2]{S} \\
    &=2 \pi \brad m_M - \pi \brad^2 \mu\ ,
\end{align}
where
$\brad$ is the radius of the monopole worldline.
Maximizing this with respect to $\brad$, we obtain
\begin{align}
    \brad&=\frac{m_M}{\mu} \,,
\end{align}
and hence
\begin{align}
    S_B^{(\mathrm{PV})}&=\frac{\pi m_M^2}{\mu}=:\pi\kappa\  \,, \label{eq:Preskill}
\end{align}
where the superscript ``(PV)'' stands for ``Preskill-Vilenkin''~\cite{Preskill:1992ck}.
Self-consistency of this thin string limit requires  $\rho_E^* $ to  be larger than the string thickness $\sim 1/\sqrt{\mu}$, implying $V\gg v$.
The string breaking rate per unit length is then given by
\begin{equation}
    \Gamma \simeq \frac{\mu}{2\pi} e^{-S_B^{(\mathrm{PV})}}\ \,,
\end{equation}
with the prefactor first computed in~\cite{Monin:2008mp}.
This decay rate is key to understanding the phenomenology of metastable cosmic strings, and in particular their GW spectrum~\cite{Buchmuller:2021mbb}. Of particular interest is the region where cosmic strings are sufficiently long-lived to produce GWs over a wide range of frequencies, while decaying before the decoupling of the CMB so as to avoid constraints from direct searches for cosmic strings in the CMB. According to the estimate above, this occurs for $\sqrt{\kappa} \sim 6 - 8$, which points to a mild hierarchy of scales, $V \gtrsim v$, and thus the need to go beyond the thin string, or decoupling, limit.

%%%%%%%%%%%%%%%%%%%%
\section{Numerical calculation of bounce solution}
\label{sec:numerics}

The goal of this section is to go beyond the thin string approximation reviewed above, which assumes hierarchical symmetry breaking scales and treats monopoles and strings as isolated objects. To this end, we will numerically evaluate the bounce action associated with spontaneous monopole formation on the string (i.e.\ string decay) using gradient flow to determine the Higgs and gauge field profile functions.

Before solving the equations of motion,
it is worth mentioning the fact that 
while the model contains six parameters ($\lambda, \tilde{\lambda}, \gamma, v, V, g$),
the action only depends on five independent dimensionless parameters (see Appendix~\ref{app:convention}):
\begin{align}
    \frac{m_{h_1}^2}{m_\gamma^2},\quad 
    \frac{m_{\phi}^2}{m_W^2}, \quad 
    \frac{m_{h_2}^2}{m_W^2} , \quad 
    \frac{m_W^2}{m_\gamma^2}, \quad
    g \, .
    \label{eq:five-parameters}
\end{align}
This allows us to adopt units of $g v = 1$ to implement the dimensionful parameters in our simulations.  We further note that $g$ does not affect any dynamics
but only changes the overall value of the action. This implies that the equations of motion depend only on the four parameters other than $g$, while we need to account for $g$ when evaluating the string decay rate.

\subsection{Ans\"atz}
We assume two symmetries:
the rotational symmetry in the $x$-$y$ plane
since the string and monopole are located on the $z$-axis,
and the Lorentz invariance in the $t$-$z$ plane,
which is equivalent to the rotational symmetry in the $t_E$-$z$ plane.
Thus it is convenient to introduce two ``polar coordinates''
\begin{align}
    x=\rho\cos \varphi,& \quad y=\rho \sin\varphi \quad  \\ 
    z= \rho_E \cos \psi,&\quad t_E(=-i t)= \rho_E \sin\psi
    \label{eq:coordinates}
\end{align}
with $\varphi,\psi\in[0,2\pi)$.
These two symmetries should be understood as imposed up to gauge transformations,
and hence the scalar fields can in principle depend on the angles $\varphi$ and $\psi$ that would be canceled by gauge fields in gauge-invariant quantities.
This is indeed the case for cosmic strings since their fields have winding phases in two-dimensional space,
which induce finite magnetic fluxes.
In the current case, we allow such a $\varphi$-dependence to describe the string on $\rho=0$ while $\psi$-dependence should not exist because it would introduce an extra string-like object located on $\rho_E=0$.
This additional assumption simplifies the gauge fields since $A_\psi=0$ solves its equation of motion.

Within these assumptions,
we work with the most general ans\"atz:
\begin{align}
    H &= v \Big[f(\rho,\rho_E) + i \,h(\rho,\rho_E) \sigma_\rho + i\, d(\rho,\rho_E) \sigma_\varphi +i \, e(\rho,\rho_E) \sigma_3\Big]
    \begin{pmatrix}
        0\\1
    \end{pmatrix} \label{eq:H-ansatz}\\
\phi &= \frac{V}{2} \big[a(\rho,\rho_E)\sigma_\rho + b(\rho,\rho_E) \sigma_\varphi + c(\rho,\rho_E)\sigma_3 \big] \\[2mm]
A_\rho &= u_1 (\rho,\rho_E) \frac{\sigma_\varphi}{2} + u_2(\rho,\rho_E) \frac{\sigma_\rho}{2} + u_3(\rho,\rho_E) \frac{\sigma_3}{2} \\[2mm]
A_\varphi &= u_4 (\rho,\rho_E) \frac{\sigma_\varphi}{2} + u_5(\rho,\rho_E) \frac{\sigma_\rho}{2} + u_6(\rho,\rho_E) \frac{\sigma_3}{2} \\[2mm]
A_{\rho_E} &= u_7 (\rho,\rho_E) \frac{\sigma_\varphi}{2} + u_8(\rho,\rho_E) \frac{\sigma_\rho}{2} + u_9(\rho,\rho_E) \frac{\sigma_3}{2}  \\[2mm]
A_\psi &=0 \, , \label{eq:Wpsi-ansatz}
\end{align}
with $\vec A = A_\rho \hat e_{\rho} + (A_\varphi/\rho) \hat e_\varphi + A_z \hat e_z$ where $\hat e_{\rho, \varphi, z}$ are the usual unit vectors in radial, polar and $z$ directions, and
\begin{align}
    \sigma_\varphi \equiv -\sin\varphi\, \sigma_1 + \cos\varphi\, \sigma_2 , \quad 
    \sigma_\rho \equiv \cos\varphi\, \sigma_1 + \sin\varphi\, \sigma_2 \, .
    \label{eq:ansatz}
\end{align}
We have 16 unknown functions in total.
\footnote{The $(0,1)^t$ vector in $H$ is just a choice of gauge. We could start with another unit vector, which is related to the above one via a gauge transformation, and obtain the equivalent result.
}

\subsection{Gradient flow with fixed monopole core}
The bounce solution is a saddle-point of the action.
This means that if one considers field-configuration space,
the bounce solution corresponds to a stationary point that extremizes the action, instead of a minimum or maximum point of the action.
Around the solution, the curvature (second functional derivative of the action) has both positive and negative values depending on the directions.
This leads to  a complication since a simple minimization/maximization procedure does not work.
However, in most bounce solutions in flat spacetime, there is only one direction with negative curvature (called negative mode),
and hence the other directions have positive curvature,
so that the bounce is the maximum point along the direction of the negative mode and is the minimum point for the other directions.
Especially, in the current case, the negative mode corresponds to the shift of the position of the monopole.\footnote{
This can be seen by explicit computation in the limit of point-like monopoles. We assume here that this result carries over to extended monopoles.
If this assumption were incorrect, our gradient flow would not converge to non-trivial configurations.
Thus the validity of the assumption is verified \textit{a posteriori} by obtaining such solutions.
}
Once the position of the monopole is fixed, the bounce is a point minimizing the action within the constrained configuration space.

With these preliminaries,
we obtain the bounce solution as follows.
We promote the functions in Eqs.~\eqref{eq:H-ansatz}-\eqref{eq:Wpsi-ansatz} to be functions of $\rho,\rho_E,\tau$
with $\tau$ being a fictitious time called the flow time.
Without loss of generality, we fix the worldline of the monopole to be
\begin{align}
    a(0,R,\tau) = b(0,R,\tau) = c(0,R,\tau) = 0 \,, \label{eq:constraint}
\end{align}
with a constant $R$. This fixes the position of the monopole ($\phi = 0$) onto the core of the string ($\rho = 0$) at $\rho_E = R$. Making use of the symmetries discussed above, we only need to study the bounce action associated with the formation of a single monopole, where it is understood that the corresponding antimonopole formation is the symmetric process in $-z$ direction.
We then solve the gradient flow equation
\begin{align}
    \partial_\tau X(\tau,\rho,\rho_E) = -\frac{\delta S}{\delta X(\tau,\rho,\rho_E)} \label{eq:gradient-flow}
\end{align}
with $X$ being the functions given in Eqs.~\eqref{eq:H-ansatz}-\eqref{eq:Wpsi-ansatz},
$X=\{a,b,c,f,h,d,e,u_1,\cdots,u_9\}$.
It is easy to show that the action decreases during the flow, $\partial_\tau S = -\sum_X \int d^4x \left(\delta S/\delta X\right)^2 \leq 0$.
Once this flow converges, $\delta S/\delta X=0$,
that configuration minimizes the action with the constraint~\eqref{eq:constraint},
whose minimum value is denoted by $S_*(R)$.
By performing the same procedure with different $R$, we obtain a function $S_*(R)$.
The bounce solution is obtained by \textit{maximizing} $S_*(R)$,
i.e.,
\begin{align}
   S_B= \underset{R}{\mathrm{max}}\, (S_*(R))\,.
\end{align}

In this method, we should prepare an appropriate ``initial configuration'' for the time evolution \eqref{eq:gradient-flow} at $\tau=0$, to ensure that the flow \eqref{eq:gradient-flow} converges.
As a good initial configuration, 
we utilize the ``primitive ans\"atz'' in Ref.~\cite{Chitose:2023dam}, which interpolates between the false and the true vacua with a tanh-like function. 
Note that we just start from this configuration at $\tau=0$
but do not impose those ans\"atze during the time evolution.
For instance, we have $u_1,u_2,u_3=0$ at $\tau=0$ ,
while we do not impose it during $\tau \neq 0$ but allow them to be non-zero.

\subsection{Boundary conditions}
We need to impose correct boundary conditions at $\rho= \{0, \infty \}$ and $\rho_E=\{0,\infty\}$ for the functions given in Eqs.~\eqref{eq:H-ansatz}-\eqref{eq:Wpsi-ansatz}.
The conditions at infinity ($\rho=\infty$ and $\rho_E=\infty$) are simply given by demanding that the action is finite.
In addition, for $\rho_E=0$, the conditions are somewhat simple since there is no $\psi$-dependence.
Thus we impose the Neumann condition so that the derivatives of all functions vanish at $\rho_E=\{0,\infty\}$ and $\rho=\infty$.

The boundary at $\rho=0$, i.e.\ at the core of the string, is more non-trivial.
For the scalar fields $H$ and $\phi$
we impose that $\varphi$-dependent parts ($h,d,a,b$) should vanish (Dirichlet condition) for regularity
while for the other parts ($f,e,c$) we require the derivatives to vanish (Neumann condition).
On the other hand, the gauge fields are complicated
because singular gauge fields do not necessarily mean a physical singularity;
all they need to satisfy is regularity \textit{up to gauge transformations}.
Thus one should consider gauge-invariant observables,
among which we pick up $|D_\mu H|^2$.
Near $\rho=0$, 
one has
\begin{align}
    |D_\mu H|^2 = \frac{v^2}{4 \rho^2}\left( e(\rho,\rho_E)^2+f(\rho,\rho_E)^2\right) \left(u_4(\rho,\rho_E)^2 +u_5(\rho,\rho_E)^2+u_6(\rho,\rho_E)^2\right) + \mathcal{O}(\rho^{-1}) \, ,
\end{align}
from which one gets the Dirichlet conditions, $u_4=u_5=u_6=0$ (i.e.\ $A_\varphi = 0$) at $\rho=0$.  Moreover, we require the vector field $\vec{A}$ to be well defined and continuous at the core of the string. For example, considering a rotation by $\pi$ around the string axis, $\varphi \mapsto \varphi + \pi$, which implies
\begin{align}
 \hat{e}_\varphi & \mapsto - \hat{e}_\varphi \,, \quad \hat{e}_\rho \mapsto - \hat{e}_\rho \,, \quad \hat{e}_z \mapsto \hat{e}_z \,,  \nonumber  \\
 \sigma_\varphi & \mapsto - \sigma_\varphi \,, \quad \sigma_\rho \mapsto - \sigma_\rho \,, \quad \sigma_z \mapsto \sigma_z \,,
\end{align}
enforces
\begin{align}
 A_\rho(\varphi + \pi) & = - A_\rho(\varphi) \quad \rightarrow u_3 = 0 \,, \nonumber \\
A_{\rho_E}(\varphi + \pi) & = A_{\rho_E}(\varphi) \quad \rightarrow u_7 = u_8 = 0 \,.
\end{align}
Summarizing, this implies Dirichlet boundary conditions at $\rho = 0$ for
\begin{align}
   a=b=h=d= u_3 = u_4 = u_5= u_6 = u_7 = u_8 &= 0  \label{eq:Dirichlet-rho} \,,
\end{align}
while we require differentiable profiles (Neumann boundary conditions) at $\rho = 0$ for all other functions,
\begin{align}
  \partial_\rho c = \partial_\rho f = \partial_\rho e = \partial_\rho u_1 = \partial_\rho u_2 = \partial_\rho u_9 &= 0 \, . \label{eq:Neumann-rho}
\end{align}

The gauge-invariant term $F_{\mu\nu}^aF^{\mu\nu}_a$ provides a consistency check of our boundary conditions. Inserting the expressions above with $u_4(0, \rho_E) = u_5(0, \rho_E) = u_6(0, \rho_E) = 0$ (and consequently $\partial_{\rho_E} u_4(0, \rho_E) = \partial_{\rho_E} u_5(0, \rho_E) = \partial_{\rho_E} u_6(0, \rho_E) = 0$) yields
\begin{align}
     F_{\mu\nu}^aF^{\mu\nu}_a(0, \rho_E) =   \left[( u_1 + \partial_{\rho} u_5)^2 + ( u_2- \partial_{\rho}u_4)^2 + (\partial_{\rho} u_6)^2 + u_7^2 + u_8^2\right]/(2 \rho^2) + {\cal O}(1/\rho), \label{eq:FF_rho=0}
\end{align}
where we have dropped the argument $(0, \rho_E)$ to ease the notation. Cancelling the divergence at ${\cal O}(1/\rho^2)$ thus requires
\begin{align}
  u_1+ \partial_\rho u_5=0,\quad u_2- \partial_\rho u_4=0,\quad \partial_\rho u_6=0 ,\quad u_7 = u_8=0\, ,
  \label{eq:implicit-rho}
\end{align}
at $\rho = 0$. We have checked explicitly that our numerical solutions, obtained by imposing the boundary conditions Eq.~\eqref{eq:Dirichlet-rho} and \eqref{eq:Neumann-rho}, fulfill these. One can explicitly check that Eq.~\eqref{eq:implicit-rho} also implies that the ${\cal O}(1/\rho)$ divergence in $F_{\mu\nu}^aF^{\mu\nu}_a$ is cancelled.

%%%%%%%%%%%%%%%%%%%%%%%%%%%%%%%%%%%%%%%%%%%%%%%%%%%%%%%%%%%%%%%%%%%%%%%%%%%%%%%%%%%%%%%%%%%%%%%%%%%%%%%%%
\section{Results}
\label{sec:results}
\subsection{Simulation results}
\begin{figure}
    \centering
    \includegraphics[width=0.5\textwidth]{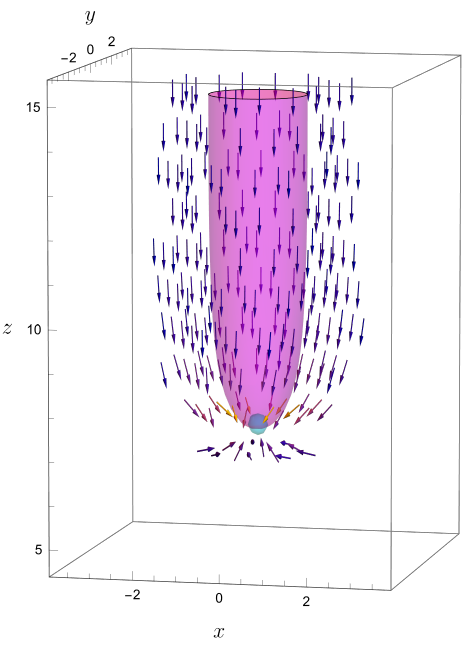}
    \caption{
    3D plot of the bounce solution,
    reconstructed for visualization from the 2D solution in the $(\rho,\rho_E)$ space at $t_E=0$.
    The magenta and cyan regions represent those satisfying $|H|^2<0.2$ and $|\phi|^2<0.2$,
    indicating the string and monopole cores, respectively.
    The arrows indicate the strength and directions of the magnetic field vector $B^{U(1)i}$.
    We have taken $m_W/m_\gamma=3\sqrt{2}$ and $\delta=0.1$ in this plot.
    In this case, the monopole-antimonopole separation at nucleation is $2 R = 3.5 m_\gamma^{-1} V/v$.
    }
    \label{fig:3Dstring-monopole}
\end{figure}

As a  benchmark case, 
we first fix three mass ratios 
$m_{h_1}/m_\gamma$, $m_\phi/m_W$, and $m_{h_2}/m_W$ to be unity
(i.e., degenerate Higgs and gauge boson masses set by the respective symmetry breaking scale)
and effectively reduce the parameter space to the 1D space spanned by $m_W/m_\gamma$.
The bounce solution obtained by our simulations is plotted in Fig.~\ref{fig:3Dstring-monopole},
in which the 2D profiles in $(\rho,\rho_E)$ space are translated into 3D $(x,y,z)$ space with $t_E=0$.
The magenta and cyan regions indicate the string and monopole cores, respectively.
The arrows indicate the magnetic field vector $B^{U(1)i}$ defined in Eq.~\eqref{eq:U(1)flux},
showing that the magnetic flux along the string is absorbed by the (anti)monopole.

\begin{figure}
    \centering
    \includegraphics[width=0.95\textwidth]{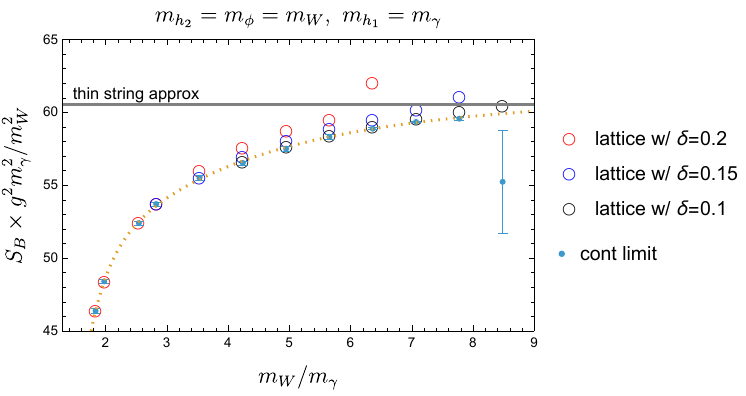}
    \caption{Results for the bounce action as a function of the ratio of the gauge boson masses, $m_W/m_\gamma = \sqrt{2} V/v$, for three different lattice spacings $\delta$ in units of $g v$. We show also the continuum limit (blue dots with $1\sigma$ error bars), which are well approximated by Eq.~\eqref{eq:fit}, shown as the orange dotted  line.
    }
    \label{fig:action-fit}
\end{figure}

The calculated bounce actions are shown as empty circles in Fig.~\ref{fig:action-fit} for different lattice spacings $\delta$ (in units of $1/(gv)$) and $m_W/m_\gamma$. 
We have performed simulations with $\delta=0.2,0.175,0.15,0.125$ and $0.1$, 
three of which are shown in Fig.~\ref{fig:action-fit} for the presentation.
As expected, the sensitivity to the lattice spacing is more significant at larger $m_W/m_\gamma$ due to the increased scale separation between the string width and the monopole profile. In this region, our results approach that obtained by the thin string approximation (gray horizontal line) as we decrease the lattice spacing, which indicates that $\delta = 0.1/(gv)$ is sufficient to capture all relevant scales in the entire regime shown. For larger values of $m_W/m_\gamma$, one can simply rely on the analytical result in the thin string limit.

%%%%%%%%%%%%%%%%%%%%
%%%%%%%%%%%%%%%%%%%%

%%%%%%%%%%%%%%%%%%%%
%%%%%%%%%%%%%%%%%%%%

This resolves the problem encountered in Ref.~\cite{Chitose:2023dam}, which found values for the bounce action significantly above the thin string limit for $m_W/m_\gamma \gg 1$. The problem was especially pronounced for the `improved ans\"atz' discussed there, which did not show the expected parametric behaviour and instead was monotonically growing over the entire regime of $m_W/m_\gamma$ considered. As the authors stated, their results provide an upper bound on the bounce action, but the deviation from the analytical result in the thin string limit indicated that the solution found was not the minimal energy one. Here, by imposing no assumptions beyond symmetry considerations on the field profiles, this problem is resolved.

The results for different lattice spacings allow us to extrapolate to the continuum limit, shown in Fig.~\ref{fig:action-fit} by the blue dots with $1\sigma$ error bars indicating the uncertainty of this extrapolation.
This is done by assuming that the deviation from the continuum limit due to the non-zero lattice spacing is proportional to $\delta^3$,
which can be checked \textit{a posteriori}.
(The lattice data with the largest spacing $\delta=0.2$ and $0.175$ are not used to obtain the continuum limit for $m_W/m_\gamma>6.3$ since they deviate from the $\delta^3$ scaling, decreasing the overall accuracy and precision of the extrapolation.)
The result for the continuum limit is well fitted by the following function
\begin{align}
    \label{eq:fit}
     S_B & = \frac{m_W^2}{g^2 m_\gamma^2} \sum_{i=0}^5 c_i \left(\frac{m_\gamma}{m_W}\right)^i  \equiv  \frac{m_W^2}{g^2 m_\gamma^2} {\cal F}(m_W/m_\gamma) \, ,
\end{align}
in which the unknown coefficients $c_i$'s are determined by a weighted fit as summarized in Table~\ref{tab:fit-ci}.
This fitted function $\mathcal{F}(m_W/m_\gamma)$ is plotted as an orange dotted curve in Fig.~\ref{fig:action-fit}.
The uncertainty of this fitting is quite small and hence not shown in the figure.
In order to assist the fit, 
here we have also included another data point
$S_B = 0$ for $m_W/m_\gamma \simeq 1$, indicating the absence of any potential barrier for monopole creation. 
This corresponds to a classical instability,
and can be calculated by an independent method,
see Appendix~\ref{app:classical-stability} and Ref.~\cite{Blasi:2026iyq}.

Eq.~\eqref{eq:fit}, together with the values for the coefficients given in Tab.~\ref{tab:fit-ci}, is a key result of this work. It provides the exact solution for the bounce action, and hence the string decay rate, for the benchmark case $m_{h_2} = m_\phi = m_W$ and $m_{h_1} = m_\gamma$, i.e.\ degenerate gauge boson and Higgs masses within each sector, respectively. 
The asymptotic value of this function $\mathcal{F}$ in the thin string limit, $m_W/m_\gamma \rightarrow \infty$, is given by the coefficient $c_0 = 61.1 \pm 2.4$.
Notably, this is consistent with the thin string limit ($\simeq 60.5$) within the uncertainty,
which quantitatively confirms the above statement that we recover the thin string limit for large~$m_W/m_\gamma$.

\begin{table}[tbp]
    \centering
    \begin{tabular}{|c|c|c|c|c|c|}
    \hline
    $c_0$ & $c_1$ & $c_2$ & $c_3$ & $c_4$ & $c_5$ \\
    \hline
       $61.1$ &  $14.8$ & $-308.7$ & $1004.3$&  $-1410.3$ & $638.9$ \\
         \hline
    \end{tabular}
    \caption{Best-fit values of $c_i$'s in Eq.~\eqref{eq:fit}.
    }
    \label{tab:fit-ci}
\end{table}

This method can straightforwardly be extended to (mild) hierarchies between the Higgs and gauge boson masses. As an example, we show in App.~\ref{app:mass-ratio} results for the case that the mass of the $U(1)$ Higgs boson is twice the gauge boson mass, $m_{h_1} = 2 m_\gamma$. We observe a qualitatively similar behaviour, approaching the thin string limit for small lattice spacings and large $m_W/m_\gamma$. Much larger hierarchies become numerically more challenging as they introduce scale separations in the lattice simulations, but we take these examples as proof of principle of our methodology which can in principle be adapted to the mass ratios found in any specific model.

\subsection{Implications for gravitational wave searches}

A network of cosmic strings evolves by continuously splitting off cosmic string loops from self-intersecting strings, which in turn decay emitting gravitational radiation. If the cosmic strings are stable (or have a lifetime longer or comparable to the age of the Universe) this maintains the string network in the scaling regime, in which a constant fraction of the energy budget of the universe is stored in the cosmic string network~\cite{Vilenkin:2000jqa}. This leads to SGWB with an approximately scale invariant spectrum over many decades in frequency~\cite{Auclair:2019wcv}. If the cosmic strings are metastable, there is an additional decay channel through monopole antimonopole formation, which leads to a suppression of the GW spectrum at low frequencies (corresponding to GWs emitted from the network at late times)~\cite{Leblond:2009fq,Buchmuller:2021mbb}. See App.~\ref{app:SGWB} for a brief review.

\begin{figure}[tbp]
\centering
 \includegraphics[width = 0.7\textwidth]{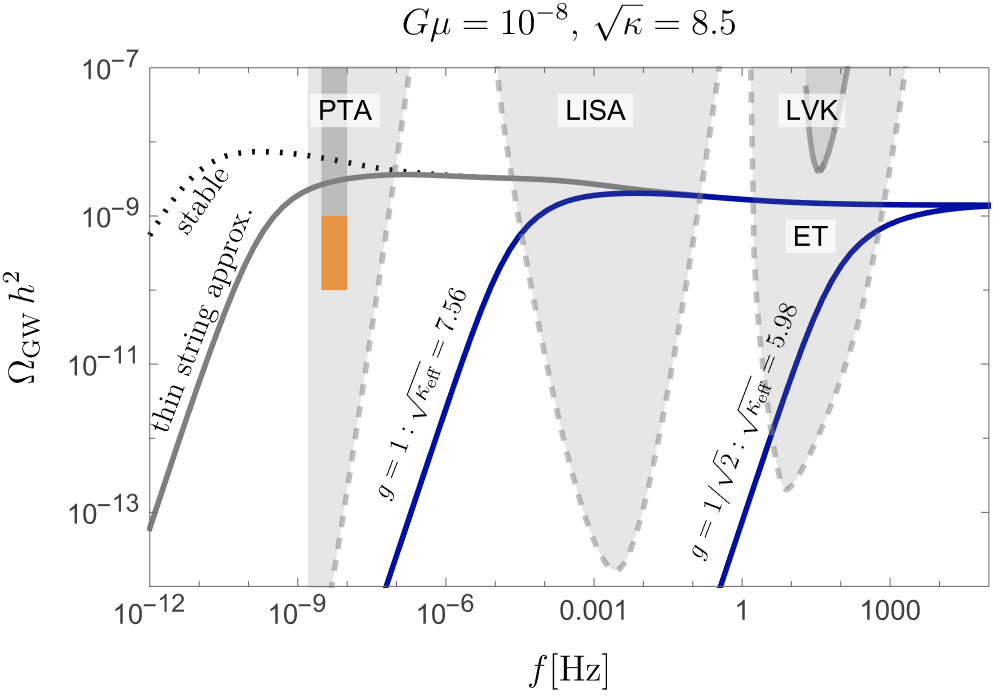}
 \caption{GW spectrum from metastable cosmic strings (dark blue) for different values of $g$, compared to the thin string approximation (solid gray) and the stable string limit (dotted black). Darker gray regions are exclusion limits, while lighter gray shaded regions show expected sensitivities. The orange box indicates the signals in pulsar timing arrays.
 }
 \label{fig:spectra_kappa_eff}
\end{figure}

The key parameters determining the GW spectrum are the string tension $\mu$ which sets the overall amplitude, and the decay rate, which is determined by the bounce action evaluated above,
\begin{align}
\Gamma \simeq \frac{\mu}{2 \pi} e^{- S_B} \,.
\end{align}
We recall that in the thin string limit,
\begin{align}
 S_B^{(PV)} = \pi \kappa = C_{PV} \frac{m_W^2}{g^2 m_\gamma^2} \,, \quad \text{with  } C_{PV} = 4 \pi^2 \frac{c_M^2}{c_S} \simeq 60.5\,,
\end{align}
where we have used $m_W = m_\phi$ and $m_\gamma = m_{h_1}$ to evaluate $C_{PV}$. As we have seen above, taking into account the finite thickness of the string reduces the bounce action,
\begin{align}
 S_B =  \frac{m_W^2}{g^2 m_\gamma^2} {\cal F}(m_W/m_\gamma) \equiv \pi \kappa_\text{eff} \,, \label{eq:kappa_eff}
\end{align}
with ${\cal F}(m_W/m_\gamma)  \leq C_{PV}$ and hence $\kappa_\text{eff} \leq \kappa$. This correspondingly enhances the string decay rate.
The earlier decay of the cosmic string network moves the turnover point below which the GW spectrum is suppressed to larger frequencies. This is illustrated in Fig.~\ref{fig:spectra_kappa_eff} for a fixed value of $G\mu = 10^{-8}$ and a fixed value of the bare parameter $\kappa = m_M^2/\mu = 8.5^2$. We show the thin string approximation (gray) as well as the result obtained using our numerical results (blue) for the bounce action~\eqref{eq:fit} (for the case $m_{h_2} = m_\phi = m_W$, $m_{h_1} = m_\gamma$) for two different values of the gauge coupling constant $g$. The impact of the finite string width on the spectrum is quite dramatic, in particular as $g$ is reduced. We note that the spectrum obtained in the thin string approximation can be recovered by adjusting the mass hierarchy $v/V$ to increase $\kappa$ accordingly. 

\begin{figure}
\centering
 \includegraphics[width = 0.48\textwidth]{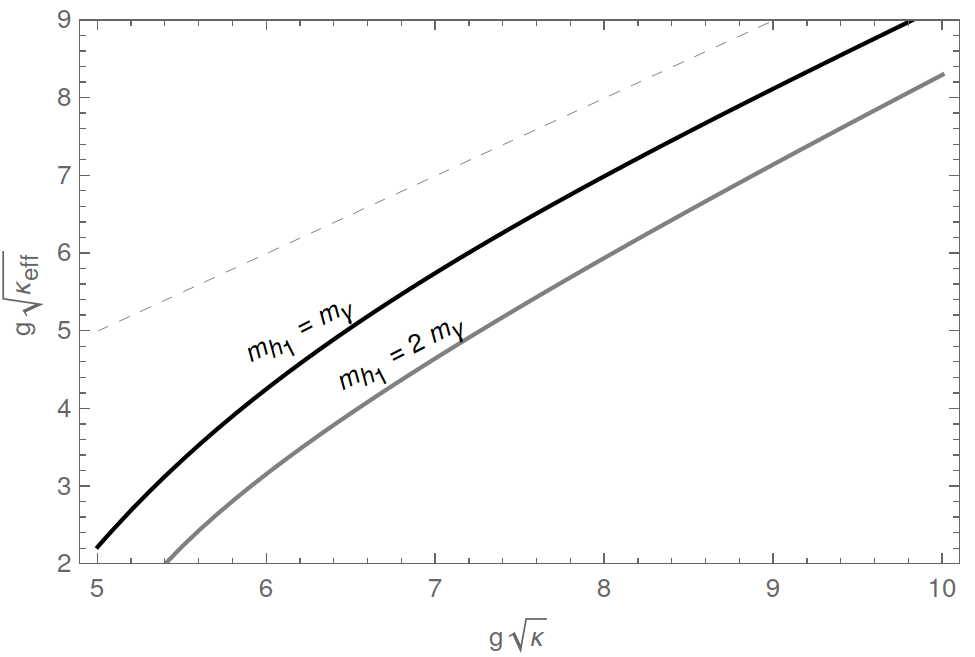} \hfill
  \includegraphics[width = 0.48\textwidth]{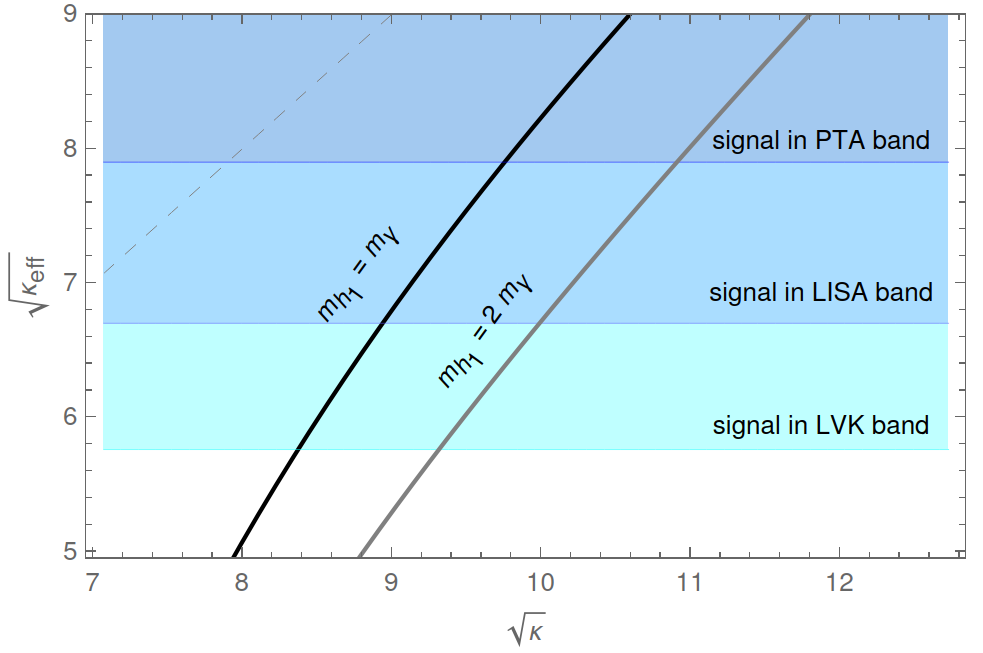}
 \caption{Correction to the string decay width due to finite width (solid black and gray), expressed by parametrizing the bounce action $S_B = \pi \kappa_\text{eff}$ in terms of the bare parameter $\kappa$, which governs the dynamics in the thin string limit. The thin string approximation is shown as dashed line. In the right panel, we have fixed  $g = 1/\sqrt{2}$ and show regions of particular phenomenological interest for GW  searches. The PTA, LISA and LVK bands are based on the low-frequency cutoff of the GW spectrum due to string decay and shown for a reference value of $G\mu = 10^{-10}$, i.e.\ neglecting the logarithmic dependence on $\mu$.
 See text and App.~\ref{app:SGWB} for details.
 }
 \label{fig:kappa}
\end{figure}

In other words, the predictions for the GW spectrum taking into account the finite string width can be obtained from the existing templates in the thin string approximation by interpreting $\kappa$ as $\kappa_\text{eff}$.
This mapping from the model parameters to $\kappa_\text{eff}$ is obtained from Eq.~\eqref{eq:fit} and illustrated in Fig.~\ref{fig:kappa}, enabling us to  translate preferred/excluded regions of the GW parameter space (in terms of $\kappa_\text{eff}$) to the bare ratio of scales contained in $\kappa$. We show results both for the benchmark discussed above, $m_{h_2} = m_\phi = m_W$, $m_{h_1} = m_\gamma$ (black)  as well as the case $m_{h_1} = 2 m_\gamma$ (gray) presented in App.~\ref{app:mass-ratio}.
As we decrease the value of $\kappa_\text{eff}$, the lifetime of the cosmic string network is decreased, moving the low-frequency cutoff of the GW spectrum to larger frequencies. This successively suppresses the signal in the pulsar timing array (PTA), LISA and LVK bands, as shown in the right panel of Fig.~\ref{fig:kappa} for  $g = 1/\sqrt{2}$ (motivated in GUTs to obtain gauge coupling unification). These results can trivially be rescaled to other values of $g$. Values of $\sqrt{\kappa_\text{eff}} \geq 9$ correspond to effectively stable strings, with a lifetime longer than the age of the universe.

\begin{figure}[tbp]
    \centering
    \includegraphics[width=0.47\textwidth]{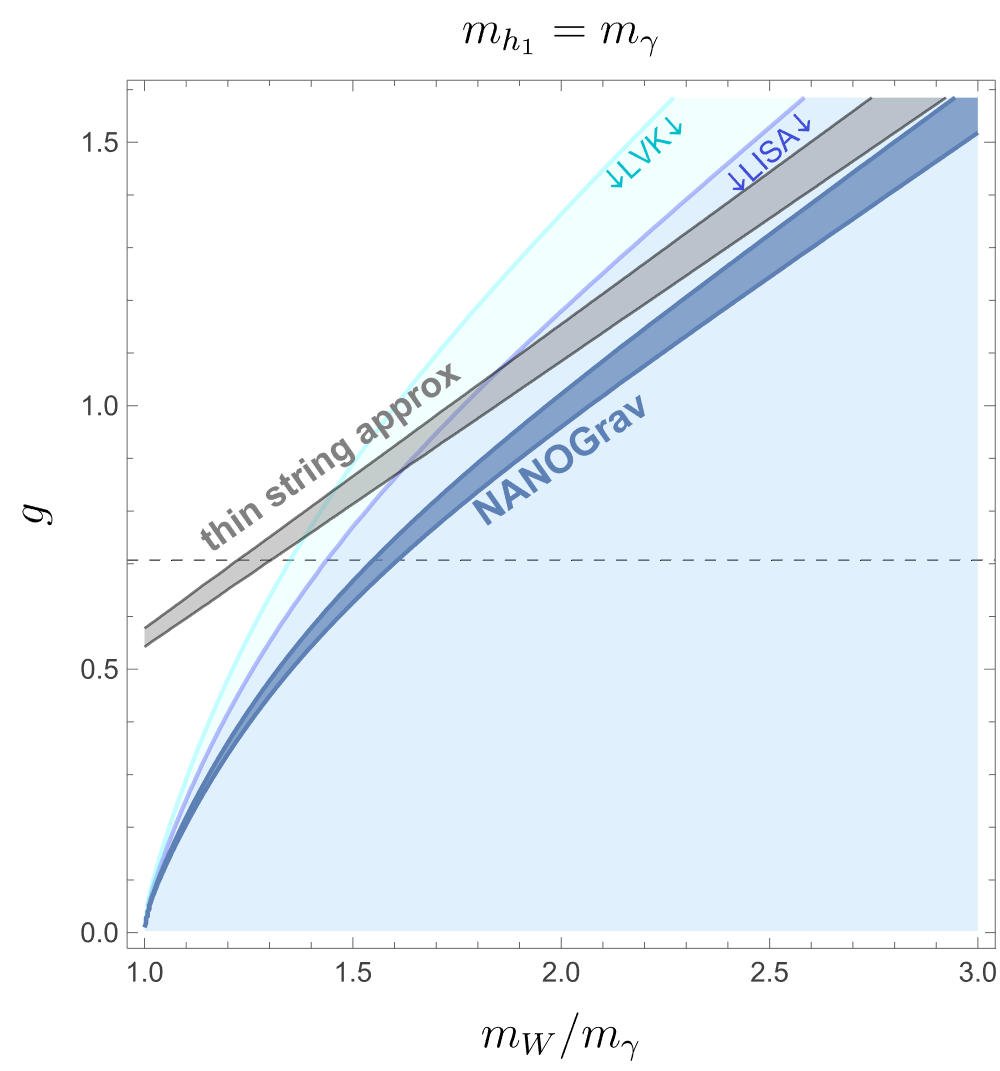}
    \hfill
    \includegraphics[width=0.47\textwidth]{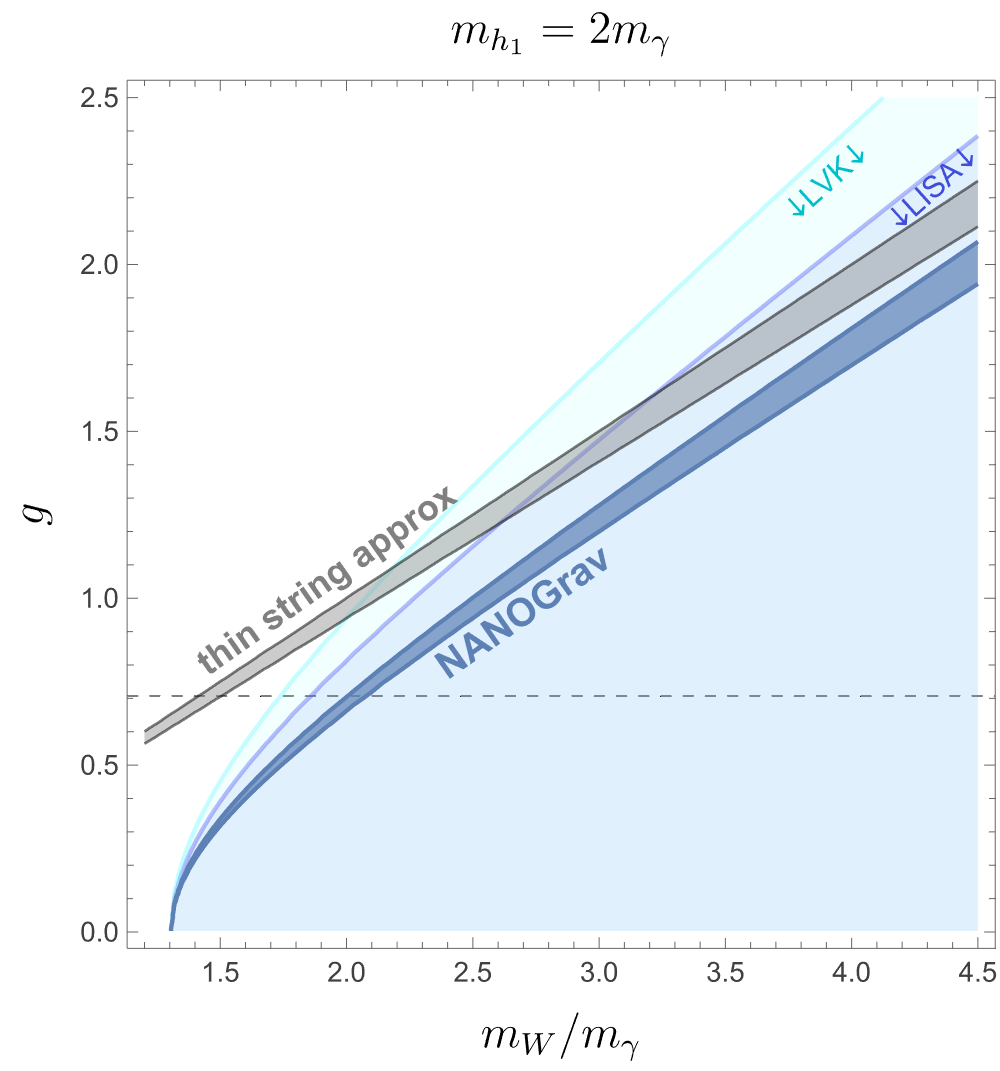}
    \caption{
    Regions of particular interest for GW phenomenology in the model parameter space. The darker shaded regions show the preferred regions when interpreting the tentative signal at NANOGrav as originating from metastable cosmic strings. The dark blue band shows the result of this work, while the gray band shows corresponding results in the thin string approximation. As in Fig.~\ref{fig:kappa}, the light blue and cyan regions indicate when the SGWB from metastable cosmic strings extends to the LISA and LVK band, respectively. The results are shown for $m_{h_1}=m_\gamma$ (left) and $m_{h_1}=2 m_\gamma$ (right) with in both cases $m_{h_2}=m_\phi=m_W$. The dashed horizontal line indicates a benchmark value $g=1/\sqrt{2}$ suggested by the gauge coupling unification.
    }
    \label{fig:parameter_space}
\end{figure}

The implications for the model parameters are illustrated in Fig.~\ref{fig:parameter_space}. For our two examples $m_{h_1} = m_\gamma$ and $m_{h_1} = 2 m_\gamma$, we show regions of interest for GW phenomenology in the $m_W/m_\gamma$-$g$ plane. The darker shaded regions show the preferred region obtained when interpreting the recent hint for an SGWB at PTAs as originating from metastable cosmic strings. These are obtained in a two-step procedure. First, we use PTArcade~\cite{Mitridate:2023oar} to infer the preferred values of the spectral parameters $\kappa_\text{eff}$ and $G\mu$ based on the NANOGrav 15-yr data set~\cite{NANOGrav:2023hde}, as shown in Fig.~\ref{fig:posterior_kappa_eff} in App.~\ref{app:SGWB}. This gives $\kappa_\mathrm{eff}=7.7$-$8.2$ after marginalizing over $G\mu$. This reproduces the results obtained in Ref.~\cite{NANOGrav:2023hvm} when interpreting $\kappa_\text{eff} = \kappa$.  Second, we use the mapping in Eq.~\eqref{eq:fit} to map the preferred regions for the posterior distributions onto the model parameters $m_W/m_\gamma$ and $g$, marginalizing over $G\mu$. This gives the dark blue shaded contour in Fig.~\ref{fig:parameter_space}  (labeled by ``NANOGrav''). In terms of $\kappa$, the best-fit value of $\sqrt{\kappa_\text{eff}} \simeq 8$ translates e.g.\ to values of $\sqrt{\kappa} \simeq 9.8$ for $g = 1/\sqrt{2}$ and $m_\gamma = m_{h_1}$. The gray shaded region is the corresponding result in the thin string approximation, i.e. for $\kappa_\text{eff} = \kappa$.  As in the right panel of Fig.~\ref{fig:kappa}, the light blue shaded regions indicate when the signal lies in the LISA (blue) and LVK (cyan) bands, for a reference value of $G\mu = 10^{-10}$.

The role of the gauge coupling in Fig.~\ref{fig:parameter_space} can be understood as follows. In the thin string approximation, $\sqrt{\kappa} = m_M/\sqrt{\mu} \propto m_W/m_\gamma/g$, i.e.\ a trivial $g$-dependence arises through the different $g$ dependence of the different mass ratios. In the full solution, the parameter rescaling discussed in App.~\ref{app:convention} makes it convenient to use $m_W/m_\gamma \propto g \sqrt{\kappa}$ as dimensionless measure of scale hierarchy. The deviation of the full solution from the thin string approximation increases for small scale separations, i.e.\ for small $\kappa$ and small $g$.
The limits $m_W/m_\gamma\to 1$ (left panel) and $1.3$ (right panel) correspond to the critical values for the classical instability, $\mathcal{F}\to 0$ in Eq.~\eqref{eq:kappa_eff}. However, when taking simultaneously the limit $g\to 0$, $S_B$ stays finite. This is why the three blue regions converge around $g=0$ and $m_W/m_\gamma=1$ or $1.3$, which is not observed in the thin string approximation since $\mathcal{F}(m_W/m_\gamma)=C_{PV}$ there.

%%%%%%%%%%%%%%%%%%%%
%%%%%%%%%%%%%%%%%%%%

As Fig.~\ref{fig:posterior_kappa_eff} in App.~\ref{app:SGWB} shows, interpreting the NANOGrav signal as originating from metastable cosmic strings leads to a preferred region in parameter space which exceeds the O4 exclusion limits from LVK~\cite{LIGOScientific:2025bgj} (independent of the microphysical origin of $\kappa_\text{eff}$).\footnote{See also the analysis of cosmic string model A in Ref.~\cite{LIGOScientific:2025kry}, which as the stable limit of the setup discussed here give similar predictions in the LVK band for parameter choices of interest to the PTA analysis.}
There are however some important caveats. First, the tentative SGWB signal observed at the different PTAs is currently undergoing scrutiny in the context of the international pulsar timing array, in particular in view of noise modeling, which may slightly shift the preferred amplitude (and hence the inferred value of $G\mu$) to lower values. Second, the assumptions for the SMBHB population may need to be revisited as astrophysical models are updated to account for the most recent data. This might allow for an explanation of the PTA signal as originating partially from SMBHs and partially from metastable cosmic strings, changing the preferred parameter values for the latter. Third, all contours rely on the prediction of the spectral tilt of the spectrum as derived in Ref.~\cite{Buchmuller:2021mbb}. While our results on the evaluation of the bounce do not affect the spectral tilt, the modifications of the breaking mechanism discussed in Refs.~\cite{Tranchedone:2026lav} and \cite{Asl:2026zpj}, do. We will return to this point below. Finally, imposing the LVK limits  assumes that the string network has already formed at around $10^8$~GeV which corresponds to the temperature scales probed by ground-based interferometers, which seems a plausible assumptions for GUT-inspired symmetry breaking processes, but nevertheless remains an assumption.
This discussion emphasizes that current multiband GW observations are probing and constraining these phenomena, providing a unique window to GUT scale physics.

Our results have interesting implications for GUT model building. Following the recipe above, the parameter space of a model in terms of couplings and ratio of scales can be mapped into the phenomenology in the thin string limit, depending only on two parameters $\kappa_\text{eff}$ and $\mu$. However, this parameter space now contains larger ratios of the bare scales, which facilitates in particular the implementation in hybrid inflation, diluting away the monopoles formed in the first phase transition before the onset of the second phase transition which forms the cosmic strings. See e.g.~\cite{Buchmuller:2019gfy,Antusch:2023zjk,Antusch:2024nqg,Antusch:2025xrs} for concrete implementations.

%%%%%%%%%%%%%%%%%%%%
\section{Conclusions}
\label{sec:conclusions}

The arguably phenomenologically most interesting regime of metastable cosmic strings arises for a mild hierarchy between the non-abelian and abelian symmetry breaking scales, as this yields cosmic strings with lifetimes which are relevant on cosmological scales and yet shorter than the lifetime of the Universe. In this regime, however, the thin string approximation, allowing for an analytical estimate of the string decay rate~\cite{Preskill:1992ck,Monin:2008mp,Leblond:2009fq}, breaks down. Building on earlier work~\cite{Chitose:2023dam}, this paper addresses this shortcoming by developing a method to numerically compute the bounce action associated with the spontaneous monopole pair creation which leads to the breaking of cosmic strings.

We solve for the field profiles of the Higgs and gauge fields involved, preserving the symmetries of the problem and implementing appropriate boundary conditions. Contrary to~\cite{Chitose:2023dam}, we impose no additional constraints on these field profiles, which allows us to find the minimal energy solution and correctly reproduce the thin string limit for hierarchical mass scales. For less hierarchical mass scales, we find a decreased bounce action which implies an increased string decay rate compared to the thin string limit. We demonstrate this for two choices of the mass spectrum of the non-abelian and abelian sector, but see no obstacles in extending our method to other choices as long as no large hierarchies within the two sectors are introduced.

Our results have implications for the target space of GUT models with observable GW spectra. Compared to the thin string approximation, our results suggest that a given string lifetime (corresponding to a low-frequency cutoff in the GW spectrum) corresponds to larger mass hierarchies between the two symmetry breaking scales. In other words, the increased string breaking rate due to finite width effects can be compensated by a larger mass hierarchy between monopole mass $m_M$ and string tension $\mu$, suppressing the spontaneous creation of the former, and recovering a similar phenomenology. This can be accounted for by a parameter $\kappa_\text{eff}$ which controls the exponential string decay rate and replaces the bare parameter $\kappa = m_M^2/\mu$. The mapping between the two is shown in Fig.~\ref{fig:kappa}. The larger hierarchy between these scales simplifies concrete implementations of cosmological metastable string formation, as it gives more space to the inflationary phases needed between monopole and string formation.

In this context, we note the recent work~\cite{Tranchedone:2026lav}, which pointed out an independent mechanism which also increases the string decay rate, due to inhomogeneities on the strings. We have not taken this into account in the results presented here, but note that this goes in the same direction as the finite width effect discussed in this paper. This suggests that the final cosmic string decay rate may be further enhanced.

Ref.~\cite{Tranchedone:2026lav} moreover pointed out the possibility of an early decay of the cosmic string network due to thermal monopole production. In the scenario which we have in mind here, with symmetry breaking processes occurring during and at the end of inflation due to the dynamics of slowly rolling scalar fields, this effect depends on the details of the reheating stage and is likely subdominant. It is however noteworthy that this process can be important in parts of the parameter space relevant for PTA searches, and in these cases, can modify the spectral tilt of the low-frequency suppression of the GW spectrum~\cite{Asl:2026zpj}.

It would moreover be interesting to extend our analysis to the case of metastable string formation in $SU(2) \times U(1) \rightarrow U(1)$, which can be seen as a minimal implementation of the breaking pattern arising e.g.\ in SO(10) GUT theories~\cite{Buchmuller:2021dtt}.

%%%%%%%%%%%%%%%%%%%%
\section*{Acknowledgments}
The authors would like to thank
Simone Blasi, Wilfried Buchm\"uller,
Akifumi Chitose, Maxime Grandjean,
Thomas Konstandin,
Andrea Mitridate and Kai Schmitz
for useful discussions. The work of Y.H. is supported by the Deutsche Forschungsgemeinschaft under Germany’s Excellence Strategy - EXC 2121 Quantum Universe - 390833306
and by the Japan Society for the Promotion of Science (JSPS) KAKENHI Grant No. JP26K17155.

%%%%%%%%%%%%%%%%%%%%
%%%% appendix %%%%
%%%%%%%%%%%%%%%%%%%%
\appendix

\section{Rescaled variables}
\label{app:convention}
Instead of using the original Lagrangian \eqref{eq:lagrangian},
it is more convenient to factor out the gauge coupling constant
by rescaling the scalar fields:
\begin{align}
\phi \to  \phi/g , \quad H \to H/g \, ,
\end{align}
which leads to a new action and potential:
\begin{equation} \label{eq:new-action}
S= \frac{1}{g^2} \int d^4x\, \left[\frac{1}{2}(D_\mu \phi^a)(D^\mu \phi^a) + (D_\mu H)^\dagger (D_\mu H) - \frac{1}{4}F^a_{\mu\nu}F^{a\mu\nu}- V_\mathrm{Higgs}^\mr{New}(\phi,H) \right]\ ,
\end{equation}
with
\begin{equation}
    V_{\mathrm{Higgs}}^\mr{New}(\phi, H )\equiv \frac{\Hlambda}{g^2} \pqty{\abs{ H }^2-g^2v^2}^2+\frac{\Vlambda}{g^2}\pqty{\phi^a\phi^a-g^2V^2}^2+ \frac{\gamma}{g^2} \abs{\pqty{\phi-\frac{g V}{2}} H }^2 \, .
\end{equation}
(Note that the covariant derivative does not contain $g$ in our notation.)
Furthermore, since the action is a dimensionless quantity,
we can take arbitrary units to measure dimensionful quantities, which does not change the action.
Thus we take the units $gv=1$ in our numerical calculation.
In other words, every dimensionful quantity is measured by $gv$.
Thus effectively we have
\begin{equation}
    V_{\mathrm{Higgs}}^\mr{New}(\phi, H )= \frac{\Hlambda}{g^2} \pqty{\abs{ H }^2-1}^2+\frac{\Vlambda}{g^2}\pqty{\phi^a\phi^a-\frac{V^2}{v^2}}^2+ \frac{\gamma}{g^2} \abs{\pqty{\phi-\frac{V}{2v}} H }^2 \, .
\end{equation}

Note that the overall prefactor $1/g^2$ in \eqref{eq:new-action} does not enter the equations of motion
while $g$ appears only through the combinations $\lambda /g^2$, $\tilde \lambda /g^2$, and $\gamma /g^2$,
which are rewritten as
\begin{align}
    \frac{\lambda}{g^2} =  \frac{m_{h_1}^2}{8m_\gamma^2},\quad 
    \frac{\tilde\lambda}{g^2} = \frac{m_{\phi}^2}{8m_W^2}, \quad 
    \frac{\gamma}{g^2} = \frac{m_{h_2}^2}{m_W^2} \, .
\end{align}
Noting $\sqrt{2}V/v = m_W / m_\gamma$,
we have five independent parameters in this model as shown in Eq.~\eqref{eq:five-parameters}.

%%%%%%%%%%%%%%%%%
\section{Classical (in)stability}
\label{app:classical-stability}
While our main interest is metastable strings, we note that there is also a classical instability in parts of the parameter space,
indicating the absence of an energy barrier separating the string and the vacuum configuration.
In order to study the classical (in)stability,
we can again perform the gradient-flow algorithm
starting from a configuration describing a straight homogeneous string at the initial flow time $\tau=0$.

\begin{figure}
    \centering
    \includegraphics[width=0.45\textwidth]{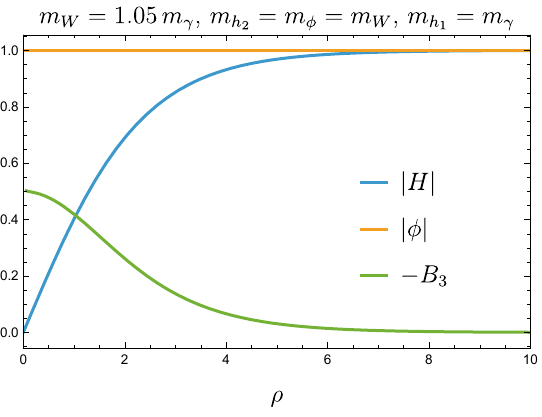}
    \hspace{2mm}
    \includegraphics[width=0.45\textwidth]{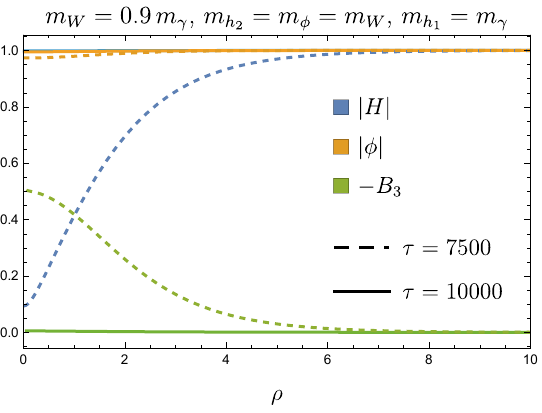}
    \caption{Classically stable (left) and unstable (right) strings.
    The stable string survives even at $\tau\to \infty$ while the unstable string starts to decay at $\tau\simeq 7500$ (dashed lines) and eventually disappears (thick lines).
    }
    \label{fig:stable_string}
\end{figure}

We impose the following restrictions on the ans\"atz Eqs.~\eqref{eq:H-ansatz}-\eqref{eq:Wpsi-ansatz}:
\begin{align}
    A_\rho=A_\psi=0  \,,
\end{align}
and $\rho_E$-independence of the other functions
(i.e., static and translationally invariant along $z$ axis),
reducing to the one-dimensional system in $\rho$ direction.
The initial string configuration is given as
\begin{align}
d(\rho,\tau=0) = \tanh(\rho), \quad c(\rho,\tau=0) =1, \quad u_6(\rho,\tau=0)=-2 \tanh(\rho)^2 \,,
\end{align}
while the other functions are set to zero (plus tiny perturbations to trigger the instability) at $\tau=0$.

If the string is classically stable, 
it is also stable against the gradient flow since it is an algorithm that decreases the static energy at every step.
The details of the initial configurations do not matter because they immediately get relaxed to be the correct string profiles.
Meanwhile, if unstable, it gradually disappears and we are left with the vacuum configuration.
Thus whether the string survives against the gradient flow tells us whether it is classically stable or not.

Let us illustrate typical string profiles evolving under the flow.
The left panel in Fig.~\ref{fig:stable_string} shows the case of a classically stable string.
One can see that there exists a string core corresponding to $|H|=0$ at $\rho=0$ and the magnetic flux $B_3$.
The right panel in Fig.~\ref{fig:stable_string} shows the case of a classically unstable string,
which disappears at the late flow time.

By repeating this with various $m_W/m_\gamma$,
one can obtain its critical value for the stability.
For instance, in the case of parameter choice in Fig.~\ref{fig:stable_string}
(same as those in Fig.~\ref{fig:action-fit}),
it turns out to be $m_W/m_\gamma=1$ as the critical one.
One also obtains
$m_W/m_\gamma=1.305$ for $m_{h_1}=2m_\gamma$
(parameter choice corresponding to Fig.~\ref{fig:SB_mh_twice}).
For more systematic and comprehensive studies on the classical instability in the same model, see Ref.~\cite{Blasi:2026iyq}.
We note in particular that the instability discussed in this appendix corresponds to a complete unwinding of the string, which proceeds (as we see in the right panel of Fig.~\ref{fig:stable_string}) through the formation of a Higgs condensate at the string core. See also Ref.~\cite{Blasi:2026iyq} for a related discussion.

%%%%%%%%%%%%%%%%%%%%%%%%
\section{Varying the mass ratio in the U(1) sector}
\label{app:mass-ratio}

\begin{figure}
\centering
 \includegraphics[width = 0.8\textwidth]{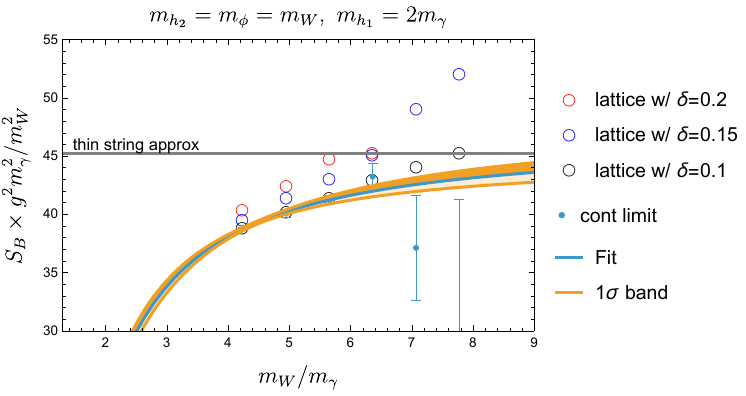}
 \caption{Same as Fig.~\ref{fig:action-fit} for a different mass ratio in the $U(1)$ sector, $m_{h_1} = 2 m_\gamma$. Compared to Fig.~\ref{fig:action-fit}, this figure is based on fewer simulations in different lattice spacings, leading to larger uncertainties in the extrapolation to the continuum limit and in fitting the coefficients of the function Eq.~\eqref{eq:fit}, as shown by the yellow band. This band would be essentially invisible for our results in Fig.~\ref{fig:action-fit}.
  }
 \label{fig:SB_mh_twice}
\end{figure}

In the main text, we focused our discussion for simplicity on the benchmark case $m_{h_2} = m_\phi = m_W$ and $m_{h_1} = m_\gamma$, i.e.\ taking the Higgs and gauge boson masses of each symmetry breaking step to be degenerate. In this appendix we show an example with a different mass ratio in the $U(1)$ sector, $m_{h_1} = 2 m_\gamma$ to illustrate a few points. Firstly, we highlight that the method presented in the main text works also in this case. Despite fewer simulations performed, we can clearly see a convergence towards the thin string limit for large $m_W/m_\gamma$ as we decrease the lattice spacing, see Fig.~\ref{fig:SB_mh_twice}. Our results are also consistently below the thin string limit, indicating that our configurations minimize the action. Second, we note that while qualitatively the results are similar to the observations in the main text, the amount of deviation from the thin string limit does depend on the mass structure of the model. This is particularly evident in Fig.~\ref{fig:kappa}, which shows the effective parameter $\kappa_\text{eff}$ controlling the GW spectrum. We observe that the case $m_{h_1} = 2 m_\gamma$ leads to a stronger suppression of the action and thus to a shorter lifetime for the cosmic strings.

%%%%%%%%%%%%%%%%%%%%%%
\section{The SGWB from metastable cosmic strings in the thin string approximation}
\label{app:SGWB}

This appendix summarizes results obtained in Refs.~\cite{Buchmuller:2019gfy,Buchmuller:2020lbh,Buchmuller:2021mbb,Buchmuller:2023aus} on the SGWB from metastable cosmic strings in the thin string approximation. In the main text, we will bootstrap these by substituting $\kappa \rightarrow \kappa_\text{eff}$ to formulate expectations for the SGWB from metastable cosmic strings beyond the thin string approximation.

\begin{figure}
\centering
 \includegraphics[width = 0.7\textwidth]{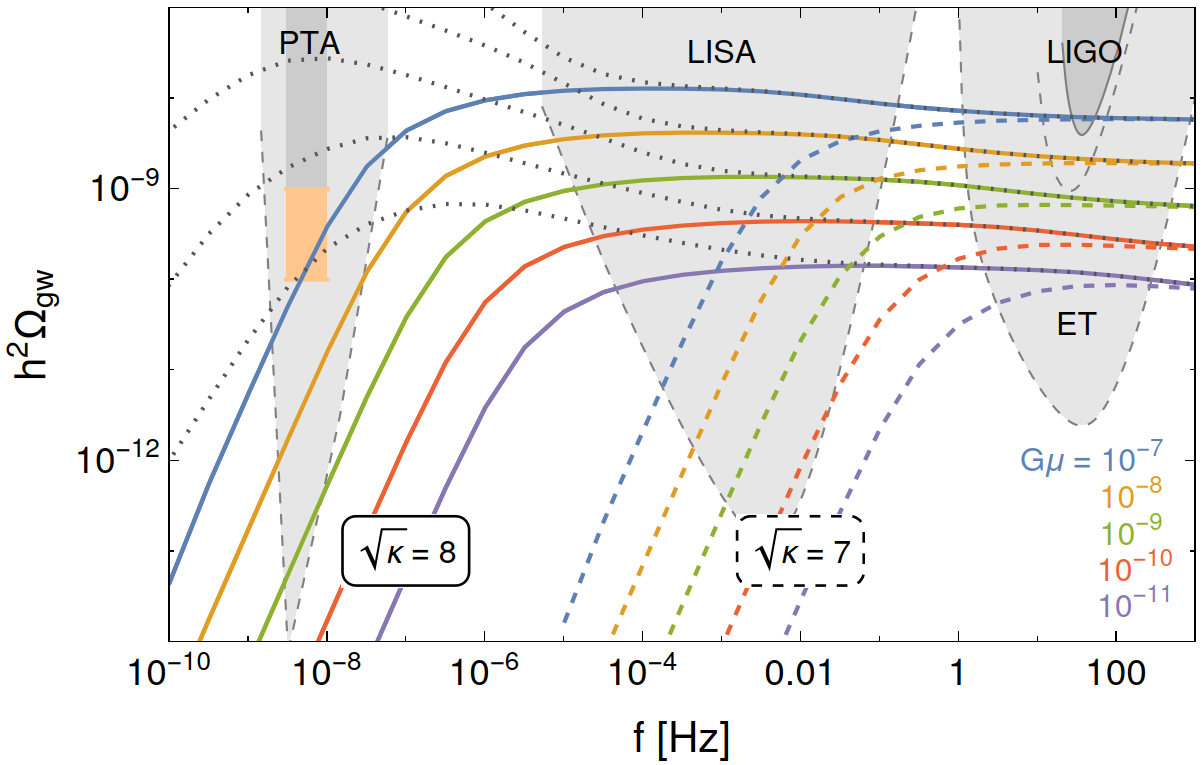}
 \caption{GW spectrum from metastable cosmic strings in the thin string approximation, for some exemplary values of $\mu$ and $\kappa$. Darker gray regions are exclusion limits, while lighter gray shaded regions show expected sensitivities. The orange box indicates the tentative signal in pulsar timing arrays. Figure adapted from \cite{Buchmuller:2021mbb}.
 }
 \label{fig:spectra}
\end{figure}

The SGWB resulting from a network of cosmic strings can be obtained in the Nambu-Goto approximation as
\begin{align}
 \Omega_\text{gw}(t_0, f) = \frac{16 \pi (G\mu)^2}{3 H_0^2 f} \sum_k k P_k \int_0^{z_i} \frac{dz'}{H(z') (1 + z')^6}\, \overset{\circ}{n}(\ell, t(z')) \,.
 \label{eq:OmegaGW}
\end{align}
Here $H(z)$ is the redshift dependent Hubble parameter, with $H_0$ its value today at $t_0$. The sum extends over the harmonic excitations of the string loops with $P_k$ the power emitted by a single loop. We integrate over all loops of length $\ell = 2 k/f(z')$ such that after red shifting, they contribute to the spectrum at frequency $f = f(0)$.
The loop number density $\overset{\circ}{n} \propto \exp(- \Gamma \ell t)$ needs to be determined by solving kinetic equations for the string network.

Compared to the spectrum obtained from stable cosmic strings, the key difference in the appearance of a scale
\begin{align}
\label{flow}
f_\text{low} \sim 3\cdot 10^{-9}\,\textrm{Hz}
\left(\frac{50}{\Gamma}\right)^{3/4}
\left(\frac{10^{-8}}{G\mu}\right)^{1/2}
\exp{\left(-\pi\left(\frac{\kappa}{4}-16\right)\right)} \,,
\end{align}
set by the decay of the cosmic string network through monopole creation. The spectrum at $f > f_\text{low}$ is sourced by GWs emitted before the decay becomes significant, and hence we recover the spectrum familiar from stable cosmic strings. In particular, for loops formed in radiation domination, we find a roughly scale invariant spectrum with an amplitude of
\begin{align}
 \Omega_\text{gw}^\text{plateau} \simeq \frac{128 \pi}{9} 0.18 \, \Omega_r \left(\frac{G \mu}{\Gamma} \right)^{1/2} \,, \label{eq:plateau}
\end{align}
where $\Omega_r h^2 = 4.15 \cdot 10^{-5}$ is the density parameter of radiation today. For $f < f_\text{low}$ the spectrum is suppressed as $f^2$, reflecting the decay of the cosmic string network. Some exemplary spectra are shown in Fig.~\ref{fig:spectra}. In $\Lambda$CDM cosmology, their shape is entirely determined by the string tension and decay rate, which in the thin string limit are parametrized by $\mu$ and $\kappa$, or equivalently $\mu$ and $m_M$.

\begin{figure}
\centering
 \includegraphics[width = 0.7\textwidth]{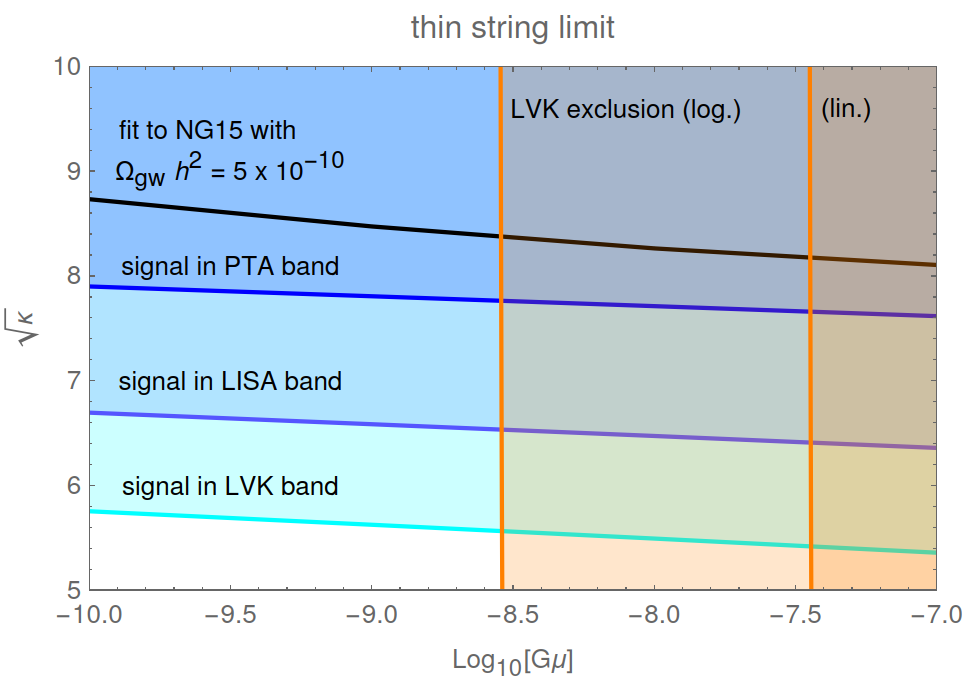}
 \caption{Regions of interest for SGWB searches from metastable cosmic strings in pulsar timing arrays, space- and ground-based interferometers in the thin string approximation.
 }
 \label{fig:SGWB-pheno}
\end{figure}

Based on this, we can determine regions in the $\kappa - \mu$ parameter space that are of particular relevance for GW searches. Searches for a scale-invariant SGWB in LIGO put an upper bound~\cite{LIGOScientific:2025bgj}
\begin{align}
 \Omega_\text{gw} (25~\text{Hz}) & \leq 2.8 \cdot 10^{-9} \quad \text{(log uniform prior)}  \\
 \Omega_\text{gw} (25~\text{Hz}) & \leq 8.6 \cdot 10^{-9} \quad \text{(linear uniform prior)}
 \,,
\end{align}
depending on the choice of prior on the GW amplitude used in the analysis.\footnote{The sensitivity curves shown in Figs.~\ref{fig:spectra_kappa_eff} and \ref{fig:spectra} are taken from \cite{LIGOScientific:2025bgj}, and are based on the linear prior.}
This implies an upper bound $G \mu < 3.0 \cdot 10^{-9}$  ($G \mu < 3.7 \cdot 10^{-8}$) for the case of the logarithmic (linear) prior.  This result was obtained from the full numerical calculation of the spectrum, but agrees well with the analytical estimate from~\eqref{eq:plateau}. The corresponding exclusion region is marked in orange in Fig.~\ref{fig:SGWB-pheno}.

Assuming that the tentative signal of an SGWB at PTAs arises from metastable cosmic strings fixes the amplitude of the SGWB at nHz scales, thus creating a one-to-one relation between $\kappa$ and $\mu$. This is shown by the black curve in Fig.~\ref{fig:SGWB-pheno} for $\Omega_\text{gw} h^2 (3 \cdot 10^{-9}\,\text{Hz}) \simeq 5 \cdot 10^{-10}$, as suggested e.g.\ by \cite{NANOGrav:2023gor}. Of course, it should be noted that there are other plausible explanations of this signal, most notably mergers of supermassive black holes, and that the best-fit value for the GW strain still comes with sizable error bars. Nevertheless, the black curve gives an indication of a particularly interesting region of the parameter space.

On the contrary, as $\kappa$ is decreased, $f_\text{low}$ crosses threshold values which leads to a strong suppression of the SGWB signal successively in the PTA, LISA and LVK bands. This is shown by the blue shaded regions in Fig.~\ref{fig:SGWB-pheno}, where we have used $f_\text{low} = 10^{-7}$, 0.1~Hz and $10^3$~Hz for estimates of the maximal frequency detectable in PTA, LISA and LVK, respectively.
These regions appear in the same color coding in Fig.~\ref{fig:kappa} in the main text, where for simplicity we have dropped the mild $G\mu$ dependence of the blue regions, and evaluated for concreteness the corresponding thresholds for $\kappa$ at $G\mu = 10^{-10}$.

%%%%%%%%%%%%%%%%%%%
\begin{figure}
    \centering
    \includegraphics[width=0.6\textwidth]{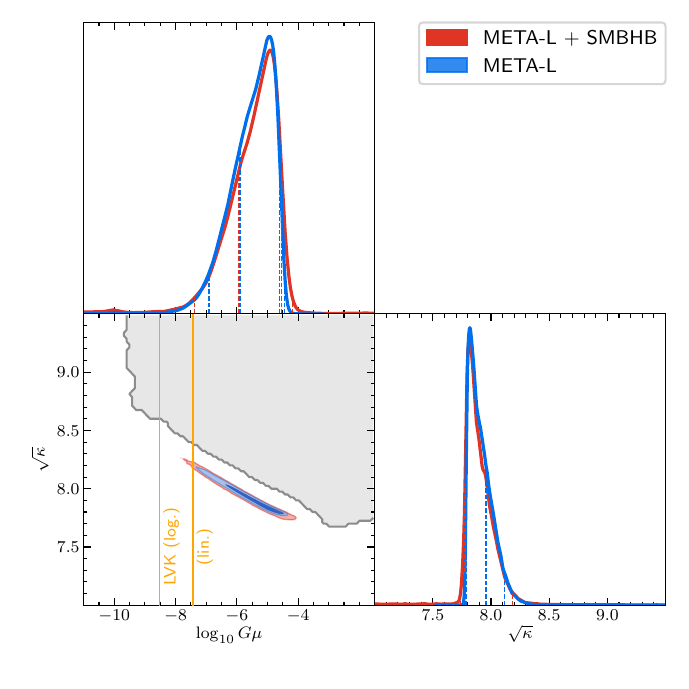}
    \caption{
    Reconstructed posterior distributions for the parameters of metastable cosmic strings for $m_{h_2}=m_\phi=m_W$, and $m_{h_1}=m_\gamma$ in the thin-string approximation.
     The blue contours show the posteriors for metastable cosmic strings as the sole source, the red contours also allow for a contribution from supermassive black holes based on \texttt{holodeck}~\cite{holodeck} (see also Ref.~\cite{NANOGrav:2023hfp}).
     On the diagonal of the corner plot, we report the 1D marginalized distributions together with the 68\% and 95\% Bayesian credible intervals (vertical lines),
     while the bottom-left panel shows the 68\% (darker) and 95\% (lighter) Bayesian credible regions in the 2D posterior distributions.
     We construct all credible intervals and regions by integrating over the regions of highest posterior density.
     The gray-shaded regions are excluded due to an overproduction of GWs. This includes also the limit of effectively stable cosmic strings at these values of $G\mu$. The region to the right of the orange lines predict a GW amplitude which exceeds the O4 LVK limit~\cite{LIGOScientific:2025bgj}. Figure generated with \texttt{PTArcade}~\cite{Mitridate:2023oar}.}
    \label{fig:posterior_kappa_eff}
\end{figure}
%%%%%%%%%%%%%%%%%%%%

The top blue region in Fig.~\ref{fig:SGWB-pheno} points to a possible region in parameter space which gives a signal in the PTA band while not exceeding the LVK limits at higher frequencies. However, when performing an actual fit to the NANOGrav 15-yr data set~\cite{NANOGrav:2023hde}, including the information on the spectral shape, we find that the O4 LVK exclusion limits rule out the entire preferred region. This is shown in Fig.~\ref{fig:posterior_kappa_eff}, which shows the posterior distributions obtained for the spectral parameters $\kappa$ and $G\mu$ from the NG15 data using \texttt{PTArcade}~\cite{Mitridate:2023oar}.  This figure reproduces well the corresponding figure in Ref.~\cite{NANOGrav:2023hvm},
%for $\kappa_\text{eff} = \kappa$.
taking into account the more recent limits from LVK.
As discussed in the main text, these results can be equally used to constrain metastable strings beyond the thin string limit by interpreting $\kappa$ as $\kappa_\text{eff}$.

\bibliographystyle{JHEP}
\bibliography{references}

\end{document}